\documentclass[preprint2]{emulateapj}

\newcommand{\EQ}{\begin{equation}}
\newcommand{\EN}{\end{equation}}

\newcommand{\Eq}[1]{eq.~(\ref{#1})}

\newcommand{\Sec}[1]{\S\ref{#1}}

\newcommand{\Fig}[1]{Figure~\ref{#1}}
\newcommand{\FFig}[1]{Figure~\ref{#1}}
\newcommand{\Tab}[1]{Table~\ref{#1}}

\newcommand{\Figss}[2]{Figs~\ref{#1}--\ref{#2}}

\newcommand{\bra}[1]{\langle #1\rangle}
\newcommand{\ff}{\mbox{\boldmath $f$} {}}
\newcommand{\kk}{\mbox{\boldmath $k$} {}}
\newcommand{\uu}{\mbox{\boldmath $u$} {}}
\newcommand{\UU}{\mbox{\boldmath $U$} {}}
\newcommand{\xx}{\mbox{\boldmath $x$} {}}
\newcommand{\oo}{\mbox{\boldmath $\omega$} {}}
\newcommand{\SSSS}{\mbox{\boldmath ${\sf S}$} {}}
\newcommand{\nab}{\mbox{\boldmath $\nabla$} {}}
\newcommand{\OO}{\mbox{\boldmath $\Omega$} {}}
\newcommand{\ii}{{\rm i}}
\newcommand{\DD}{{\rm D} {}}
\newcommand{\dd}{{\rm d} {}}
\newcommand{\K}{\,{\rm K}}
\newcommand{\g}{\,{\rm g}}
\newcommand{\s}{\,{\rm s}}
\newcommand{\cm}{\,{\rm cm}}
\newcommand{\km}{\,{\rm km}}
\newcommand{\kms}{\,{\rm km/s}}
\newcommand{\pc}{\,{\rm pc}}
\newcommand{\kpc}{\,{\rm kpc}}

\newcommand{\Myr}{\,{\rm Myr}}
\newcommand{\Gyr}{\,{\rm Gyr}}
\newcommand{\erg}{\,{\rm erg}}
\newcommand{\dyn}{\,{\rm dyn}}
\newcommand{\yapj}[3]{ #1, {ApJ,} {#2}, #3}

\newcommand{\yapjs}[3]{ #1, {ApJS,} {#2}, #3}

\newcommand{\yana}[3]{ #1, {A\&A,} {#2}, #3}

\newcommand{\yjfm}[3]{ #1, {JFM,} {#2}, #3}

\newcommand{\yaraa}[3]{ #1, {ARA\&A,} {#2}, #3}

\newcommand{\yprl}[3]{ #1, {PRL,} {#2}, #3}

\newcommand{\ymn}[3]{ #1, {MNRAS,} {#2}, #3}

\newcommand{\ypre}[3]{ #1, {PRE,} {#2}, #3}

\newcommand{\yjour}[4]{ #1, {#2}, {#3}, #4}

\newcommand{\ssapj}[1]{ #1, {ApJ}}

\submitted{Received April 11, 2006, accepted September 7, 2006}
\begin{document}
\title{Thermal instability in shearing and periodic turbulence}
\author{
Axel Brandenburg\altaffilmark{1},
Maarit J.\ Korpi\altaffilmark{2},
and
Antony J.\ Mee \altaffilmark{3}
}
\altaffiltext{1}{NORDITA, Copenhagen, Denmark and
AlbaNova University Center, Stockholm, Sweden}
\altaffiltext{2}{Observatory, University of Helsinki, Finland}
\altaffiltext{3}{School of Mathematics and Statistics, University of Newcastle,
  Newcastle upon Tyne, UK
    \\ $ $Revision: 1.116 $ $ (\today)
}

\begin{abstract}
The thermal instability with a piecewise power law cooling function is
investigated using one- and three-dimensional simulations with periodic
and shearing-periodic boundary conditions in the presence of
constant thermal diffusion and kinematic viscosity coefficients.
Consistent with earlier findings,
the flow behavior depends on the average density, $\bra{\rho}$.
When $\bra{\rho}$ is in the range $(1$--$5)\times10^{-24}\g\cm^{-3}$
the system is unstable and segregates into cool and warm phases
with temperatures of roughly $100$ and $10^4\K$, respectively.
However, in all cases the resulting average pressure $\bra{p}$ is independent
of $\bra{\rho}$ and just a little above the minimum value.
For a constant heating rate of $0.015\erg\g^{-1}\s^{-1}$,
the mean pressure is around $24\times10^{-14}\dyn$ (corresponding to
$p/k_{\rm B}\approx1750\K\cm^{-3}$).
Cool patches tend to coalesce into bigger ones.
In all cases investigated there is no sustained turbulence,
which is in agreement with earlier results.
Simulations in which turbulence is driven by a body force show that when
rms velocities of between 10 and 30 km/s are obtained,
the resulting dissipation rates rates are comparable to
the thermal energy input rate.
The resulting mean pressures are then about $30\times10^{-14}\dyn$,
corresponding to $p/k_{\rm B}\approx2170\K\cm^{-3}$.
This is comparable to the value expected for the Galaxy.
Differential rotation tends to make the flow two-dimensional, that is,
uniform in the streamwise direction, but this does not lead to
instability.
\end{abstract}


\section{Introduction}

The importance of thermal instability (TI) has been
extensively studied in the context of the generation and regulation of
structures in the atomic interstellar medium (the so-called cold and
warm atomic phases usually denoted as CNM and WNM), for a review see
e.g.\ Cox (2005).
The core idea was presented by Field et al.\ (1969,
hereafter FGH) in their famous two-phase model: two thermally stable
phases (cold and cloudy; warm and diffuse) co-exist in pressure
equilibrium regulated by the presence of a thermally unstable phase at
an intermediate temperature. After the observational determination of the
existence of significant amounts of hot gas in the Galaxy, the FGH
model was complemented with a third, hot, phase by McKee \& Ostriker
(1977), in which model most of interstellar space was occupied by
million-kelvin gas produced in supernova explosions. Since then, the
estimates of the filling factor of the hot component have reduced to
10\%--30\% near the Galactic midplane, being larger at larger
heights. Moreover, most of the hot gas seems to be confined in large
bubbles created by clustered supernova activity rather being distributed
homogeneously around the Galaxy. In this light, therefore, it seems
justified to neglect the hot component and return to the simpler
FGH picture when modeling the colder and denser phases of the
interstellar medium (ISM).

There have been a number of numerical investigations
of the interaction of turbulence and TI.
In most papers the
turbulence is forced by sources other than the TI itself: random
turbulent forcing at varying scales and Mach numbers (e.g.\ Gazol et
al.\ 2005), localized injections of energy mimicking stellar winds
(e.g.\ V{\'a}zquez-Semadeni et al.\ 2000), the magnetorotational instability
(e.g.\ Piontek \& Ostriker 2005), and systematic large-scale motions
such as propagating shock fronts (e.g.\ Koyama \& Inutsuka 2002) and
converging flows (Audit \& Hennebelle 2005) have been considered. One
of the major findings from these models is that, because of the turbulence
present in the system, large pressure deviations are generated and
significant amounts of gas can exist in the thermally unstable
regime. These results suggest that the FGH picture of the ISM
exhibiting ``discrete" temperatures and densities and a unique
equilibrium pressure should be modified in the direction of a
``continuum" of states with an overall pressure balance but with large
deviations from it.

In recent years the possibility of driving turbulence by the TI itself
has received some revived attention.
Contrary to Kritsuk \& Norman (2002a), who found turbulence to die out
as a power law, Koyama \& Inutsuka (2006) find the turbulence to be
sustained---at least for times up to $0.1\Gyr$.
The possibility of TI-induced turbulence
is potentially similar to the Jeans instability in a
 self-gravitating medium that is able to maintain a statistically
 steady state in which the instability drives the turbulence and
 turbulent heating prevents the disk from cooling into a static
 equilibrium.  Simulations by Gammie (2001) have shown that such a
 state of self-sustained ``gravito-turbulence'' is indeed
 possible. Wada et al.\ (2000) found a similar result for the case of
 the combined action of gravitational and thermal instabilities.  The
 possibility of driving turbulence by means of instabilities is indeed quite
 common in astrophysics.  Especially popular is the magneto-rotational
 instability that is known to drive turbulence in disks (Hawley et
 al.\ 1995, Brandenburg et al.\ 1995), but there is also the
 Rayleigh-Benard instability, which leads to turbulent convection (e.g.,
 Kerr 1996).

In this paper we focus on the interaction between turbulence and the
nonlinear stages of the TI,
starting from one-dimensional calculations and extending
them to three dimensions. Following an approach similar to those of Koyama \&
Inutsuka (2004) and Piontek \& Ostriker (2005), we include thermal
conduction, which stabilizes the gas at wavelengths smaller than the critical
wavenumber of the condensation mode (Field 1965).
This wavelength is usually referred to as the the Field
length; it allows the structures generated by the TI to be resolved
by the chosen numerical grid. Other approaches have also been used: In
the model of S{\'a}nchez-Salcedo et al.\ (2002), a nonuniform grid was used
to resolve all the scales down to the cooling length, but nevertheless
the required amount of gridpoints restricted the calculations to one
dimension. In some models (e.g., Gazol et al.\ 2005), no bulk viscosity
or thermal conduction is used, but they are replaced by local
resolution-dependent artificial viscosities damping Nyquist-scale
unresolvable structures. 

\section{Model}

\subsection{Governing equations}

We consider the governing equations for a compressible perfect gas,
\EQ
{\DD\ln\rho\over\DD t}=-\nab\cdot\uu,
\EN
\EQ
\rho{\DD\uu\over\DD t}=-\nab p+\nab\cdot(2\nu\rho\SSSS),
\label{dudt}
\EN
\EQ
T{\DD s\over\DD t}=2\nu\SSSS^2+{1\over\rho}\nab\cdot\left(c_p\rho\chi\nab T\right)-{\cal L},
\EN
where $\uu$ is the velocity, $\rho$ is the density, $s$ is the specific entropy,
with ${\sf S}_{ij}=\frac{1}{2}(u_{i,j}+u_{j,i})
-\frac{1}{3} \delta_{ij}\nab\cdot\uu$ being the traceless rate of strain tensor,
$\nu$ is the kinematic viscosity, $\chi$ is the thermal diffusivity,
and ${\cal L}$ is the net cooling or heating, that is,
the difference between cooling
and heating functions, with
\EQ
{\cal L}=\rho\Lambda-\Gamma,
\EN
where $\Gamma=\mbox{const}$ is assumed for the heating function. Here
we consider the photoelectric heating by interstellar grains caused by
the stellar UV radiation field, for which Wolfire et al.\ (1995) give
a value of $0.015\erg\g^{-1}\s^{-1}$ at $n=1\cm^{-3}$.

Following common practice, we adopt a perfect gas for which $\rho$ and
$s$ are related to pressure $p$ and temperature $T$ by the relations
\begin{equation}
p={{\cal R}\over\mu}\rho T,\quad
s=c_v\ln p-c_p\ln\rho+s_0,
\label{eos}
\end{equation}
where ${\cal R}=8.314\times10^7\cm^2\s^{-2}\K^{-1}$ is the
universal gas constant, $\mu$ is the mean molecular weight
(here we assume $\mu=0.62$ in all cases and neglect the effects of
partial ionization),
and ${\cal R}/\mu=c_p-c_v$, with $c_p$ and $c_v$ being the specific
heats at constant pressure and volume, respectively; $\gamma=c_p/c_v=5/3$
is their assumed ratio. The adiabatic sound speed $c_{\rm s}$ and the
temperature are related to the other quantities via
$c_{\rm s}^2=\gamma{\cal R}T/\mu$.
The specific entropy is defined up to a constant $s_0$, whose value
is unimportant for the dynamics.

We adopt a parameterization of the cooling function approximately equal to that given by
S{\'a}nchez-Salcedo et al.\ (2002), which has been obtained by fitting a
piecewise power law function of the form
\EQ
\Lambda(T)=C_{i,i+1}T^{\beta_{i,i+1}}
\quad\mbox{for $T_i\leq T< T_{i+1}$},
\label{CoolingCurve}
\EN
to the equilibrium pressure curve of the standard model of Wolfire et
al.\ (1995) for the ISM in the solar neighborhood. When thermal
equilibrium with the chosen background heating function,
$\Gamma$ is assumed, and a continuity requirement for $\Lambda$,
\EQ
C_{i-1,i}=C_{i,i+1}T_i^{(\beta_{i,i+1}-\beta_{i-1,i})},
\EN
is taken into account, we arrive at the values of the
coefficients listed in \Tab{Tcooling}.
The coefficients $C_{i,i+1}$ given by S{\'a}nchez-Salcedo et al.\ (2002)
deviate from this condition by 4\%--8\%.
It turned out that with their original coefficients the flow amplitude
showed spurious oscillations in time which disappeared when we use the
revised coefficients.

\begin{table}[t!]\caption{
Coefficients for the cooling curve given by \Eq{CoolingCurve}.
}\vspace{12pt}\centerline{\begin{tabular}{crcr}
$i$ & $T_i$ & $C_{i,i+1}$ & $\beta_{i,i+1}$ \\
\hline
1 &    10  & $3.70\times10^{16}$ & 2.12 \\
2 &   141  & $9.46\times10^{18}$ & 1.00 \\
3 &   313  &$1.185\times10^{20}$ & 0.56 \\
4 &  6102  &   $2\times10^{8}$   & 3.67 \\
5 & $10^5$ & $7.96\times10^{29}$ &$-0.65$\\
\label{Tcooling}\end{tabular}}\end{table}

It is convenient to measure time in gigayears,
speed in kilometers per second, and density in units of $10^{-24}\g\cm^{-3}$.
Pressure is therefore measured in units of $10^{-14}\dyn$.
Our unit of length is therefore $1(\kms)\times1\Gyr=1.02\kpc$;
in the following, we denote the unit of length for simplicity
as $1\kpc$, keeping in mind that it should really be $1.02\kpc$.
Viscosity and thermal diffusivity are measured in units of
$\Gyr\,\km^2\s^{-2}$.

We use periodic boundary conditions in all three directions for a
computational domain of size $(200\pc)^3$,
which is the typical domain size employed in simulations of
supernova-driven turbulence in the interstellar medium.
However, smaller domains would be more suitable to resolve the smaller
scales, as has been done by Kritsuk \& Norman (2002a), for example.
We use the \textsc{Pencil Code},\footnote{\url{see
http://www.nordita.dk/software/pencil-code}.} which is a non-conservative,
high-order, finite-difference code (sixth order in space and
third order in time) for solving the compressible hydrodynamic equations.
Because of the non-conservative nature of the code, diagnostics giving the total
mass and total energy (accounting for heating/cooling terms) are monitored
and simulations are only deemed useful if these quantities are in fact
conserved to reasonable precision.
The mesh spacings in the three directions are assumed to be the same,
that is, $\delta x=\delta y=\delta z$.

We emphasize that no shock or hyperviscosity has been
used in the present simulation.  Therefore, the only means of
stabilizing the code is through regular viscosity $\nu$ and thermal
diffusivity $\chi$.  In order to damp unresolved ripples at the mesh
scale $\delta x$ in a trail of structures moving at speed $U$, the
minimum viscosity and minimum diffusion must be on the order of
$0.01\,U\delta x$ (see Brandenburg \& Dobler 2002).
In all our simulations the velocities are subsonic, so the fastest
pattern speed is given by the sound speed.
In the following we quote the mesh Reynolds number based on the
mean (volume averaged) sound speed, $\overline{c}_{\rm s}$, and
the mesh size $\delta x$,
\EQ
\mbox{Re}_{\rm mesh}=\overline{c}_{\rm s}\delta x/\nu,
\EN
The minimum viscosity quoted above corresponds to a largest permissible
value of $\mbox{Re}_{\rm mesh}$ of about 100.
However, in the presence of strong converging flows and shocks the largest
permissible value may be of order unity.

Since we want to use
minimal values for $\nu$ and $\chi$ in both the warm and cold
components we keep $\nu$ and $\chi$ constant rather than, for example,
the dynamical viscosity or the quantity ${\cal K}\equiv\rho\gamma\chi$
(see, e.g., Piontek \& Ostriker 2004).  In the latter case, $\chi$
would vary by 2 orders of magnitude between warm and cold
phases.
If the mesh were sufficiently fine, one could allow for a physically
motivated dependence of $\chi$ on $T$, but this is neglected here.

In the calculations, we have adopted two different values of 
$\nu$ and $\chi$
($5 \times 10^{-3}$ and $5 \times 10^{-4}\Gyr\km^2\s^{-2}$),
keeping their ratio, the Prandtl number $\mbox{Pr}=\nu/\chi$, fixed to unity.
The corresponding Field lengths, calculated from  equation (\ref{kf}) using the
initial cooling timescale of approximately $1\Myr$, are
24 and $7.7\pc$, respectively.
Compared with the average value of the thermal diffusion in the neutral
ISM, roughly $6\times 10^{20}\cm^2\s^{-1}\approx2\times10^{-6}\Gyr\km^2\s^{-2}$,
corresponding to a Field length of about $0.5\pc$,
the adopted values are larger by 2--4 orders of magnitude.
The cooling length $l_{\rm cool} \approx \tau_{\rm cool} u_{\rm rms}$ is close to the
physical Field length, being roughly $0.4\pc$.
Our chosen values of $\chi$ due to the preference of a large domain size, 
are therefore too large to resolve the fine structure
in the accretion fronts that result from the cooling process.
This is a similar setup to the one investigated by Piontek \& Ostriker (2004, 2005);
models achieving Field lengths smaller than the cooling length include
for example, S{\'a}nchez-Salcedo et al.\ (2001), Kritsuk \& Norman (2004),
and Koyama \& Inutsuka (2004).
 
\subsection{Stability properties}

The first thorough stability analysis was done by Field (1965), who
also included the stabilizing effect of thermal diffusion.
Assuming the solutions to be proportional to $\exp(nt+\ii\kk\cdot\xx)$,
the dispersion relation can be written in the form
\EQ
n(n+n_\nu)(n+\beta n_\rho+n_\chi)
+\omega_{\rm ac}^2\left[n+{(\beta-1)n_\rho+n_\chi\over\gamma}\right]=0,
\label{DisperRel}
\EN
where we have also included the effect of kinematic viscosity.
Here, $\omega_{\rm ac}=c_{\rm s}k$ is the acoustic frequency
and $\beta=\dd\ln\Lambda/\dd\ln T$ is the local
logarithmic slope of the cooling function.
We have restricted ourselves to cases where $\Gamma$ is constant
and $\Lambda$ depends only on $T$.
The cooling time is characterized by the quantity
\EQ
n_\rho=\rho_0{\cal L}_\rho/(c_vT),
\EN
which is to be evaluated for the equilibrium solution.
Here, ${\cal L}_\rho=(\partial{\cal L}/\partial\rho)_T=\Lambda$.
Note that $n_\rho$ is just the inverse cooling time defined by
Piontek \& Ostriker (2004).
The subscript $\rho$ follows from a similar notation used by Field (1965)
who defined instead a wavenumber $k_\rho=n_\rho/c_{\rm s}$, which is also
referred to as the cooling wavenumber.
Viscous and diffusive effects are characterized by the corresponding rates,
\EQ
n_\nu={\textstyle{4\over3}}\nu k^2,\quad
n_\chi=\gamma\chi k^2.
\EN
Thermal instability is only possible for $\beta<1$.
This condition corresponds to the isobaric instability criterion
of Field (1965).
The isochoric and isentropic criteria, $\beta<0$ and
$\beta<-1/(\gamma-1)=-3/2$, respectively, are less strict in that the
isobaric criterion for instability is then automatically satisfied.

\begin{figure}[t!]\begin{center}
\plotone{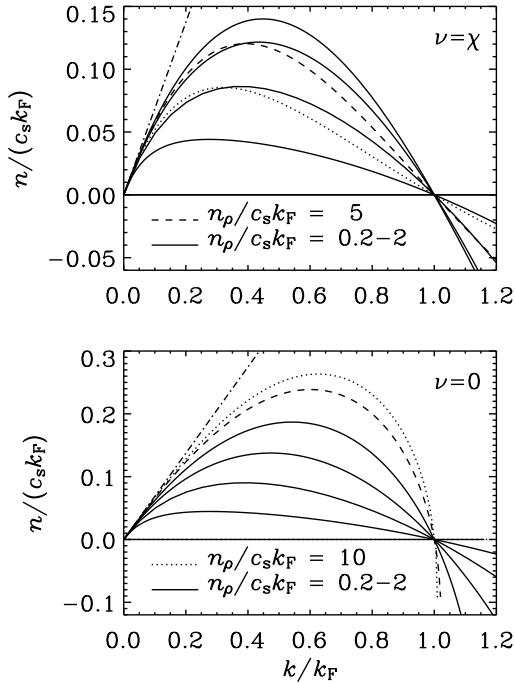}
\end{center}\caption[]{
Dispersion relation $n(k)$ in the unstable regime with $\beta=0.56$,
obtained by solving \Eq{DisperRel} for a representative range of values
of $n_\rho/(c_{\rm s}k_{\rm F})$.
The $n(k)$ curves are normalized in terms of
$k_{\rm F}$ and $c_{\rm s}k_{\rm F}$.
In the range $n_\rho/(c_{\rm s}k_{\rm F})=0.2$--2 (solid lines) the
maxima of $n/(c_{\rm s}k_{\rm F})$ are monotonically increasing.
The curves for $n_\rho/(c_{\rm s}k_{\rm F})=5$ (dashed line) and
$n_\rho/(c_{\rm s}k_{\rm F})=10$ (dotted line) deviate from this trend.
The diagonal dash-dotted line indicates the approximation
valid for small wavenumbers [\Eq{approx}].
}\label{ppdisper}\end{figure}

When thermal diffusivity is included, the gas can be stabilized
(even though $\beta<1$) provided the largest possible wavenumber
in the system (which we denote as $k_1$)
is larger than the Field wavenumber, $k_{\rm F}$, defined as
\EQ \label{kf}
k_{\rm F}^2=(1-\beta)n_\rho/(\gamma\chi)
\quad\quad\mbox{(for $\beta<1$)}.
\EN
The instability has therefore the character of an ordinary
long-wave instability requiring $k_1<k_{\rm F}$.
The corresponding dispersion relation is shown in \Fig{ppdisper}
for various values of $n_\rho$ using $\nu=\chi$ (top) and
$\nu=0$ (bottom).
The value of $\chi$ is given in terms of the ratio
$n_\rho/(c_{\rm s}k_{\rm F})$, for which three values
have been chosen to illustrate this dependence.
As expected, the presence of kinematic viscosity has a stabilizing effect.
Setting $\nu=0$ leads to somewhat larger growth rates, especially when
$n_\rho/(c_{\rm s}k_{\rm F})$ is large and $n/(c_{\rm s}k_{\rm F})$ is small.
For $n_\rho/(c_{\rm s}k_{\rm F})=2$, for example, the normalized
growth rate for $\nu=\chi$ is the largest among the three cases shown
in \Fig{ppdisper}, but it hardly increases when $\nu\to0$, in which case
the growth rate is actually the smallest among the three cases.

In the limit $k\ll k_{\rm F}$, which is relevant when diffusive effects
are negligible, the unstable branch of the dispersion relation reduces to
\EQ
n=c_{\rm iso}k\,\sqrt{\beta^{-1}-1},
\label{approx}
\EN
where we have introduced the isothermal sound speed,
$c_{\rm iso}\equiv c_{\rm s}/\sqrt{\gamma}$.
Note that in the thermally stable case with $\beta\gg1$
we obtain the usual dispersion relation for isothermal sound waves,
$\omega=c_{\rm iso}k$, where $\omega=\ii n$ is the frequency.
For $\beta=1/2$ this approximation yields $n=c_{\rm iso}k$,
as stated by Field (1965) in his equation~(36).
For $\beta=0.56$ this approximation is shown in \Fig{ppdisper} as a
straight dash-dotted line.

\subsection{Saturation properties}
\label{SaturationProperties}

\begin{figure}[t!]\begin{center}
\plotone{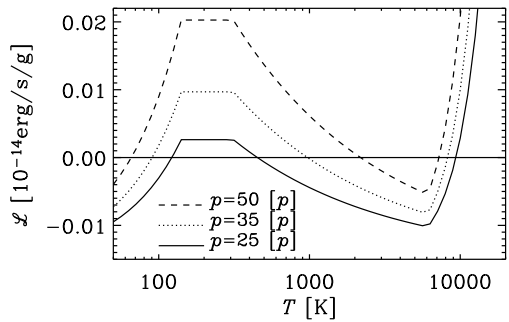}
\end{center}\caption[]{
Net cooling vs.\ temperature for three values of $p$,
given in units of $[p]=10^{-14}\dyn$.
}\label{pcoolheat}\end{figure}

\begin{figure}[t!]\begin{center}
\plotone{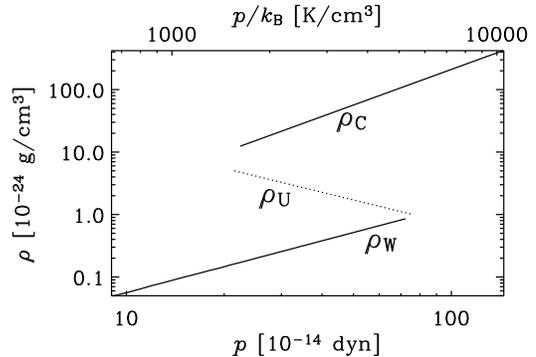}
\end{center}\caption[]{
The two stable solution branches, $\rho_{\rm C}$ and $\rho_{\rm W}$
(solid lines), and the unstable solution branch, $\rho_{\rm U}$ (dotted
line), as a function of $p$.
On the top the pressure is normalized by the Boltzmann constant,
$p/k_{\rm B}$.
}\label{ppcoolheati}\end{figure}

In the absence of thermal diffusion, thermal equilibrium is given
by the condition ${\cal L}=0$.
Pressure equilibrium between the cold and warm phases requires that
equilibrium is achieved under the constraint of constant pressure.
Such an equilibrium would however only be stable if a temperature
increase would lead to correspondingly more cooling, that is, if
\EQ
\left({\partial{\cal L}\over\partial T}\right)_p>0,
\quad\mbox{(stability)}.
\EN
where the subscript $p$ indicates that the pressure is held constant.
In \Fig{pcoolheat} we plot ${\cal L}$ as a function of $T$
for constant $p$; three values are considered: $p=25$, 35, and 50,
all in units of $[p]\equiv10^{-14}\dyn$.
This figure shows that there can be two stable states at about
$10^2$ and $10^4\K$.
We denote these values by $T_{\rm C}$ and $T_{\rm W}$ for the
cold and warm phases.
At $T\approx10^3\K$ there is an unstable equilibrium, whose temperature
is denoted $T_{\rm U}$.
The densities of the three equilibria,
obtained by solving ${\cal L}(T;p)=0$ for $T$ numerically for given $p$
and then expressing the result in terms of $\rho=\rho(T,p)$, are plotted in
\Fig{ppcoolheati}.

\begin{figure}[t!]\begin{center}
\plotone{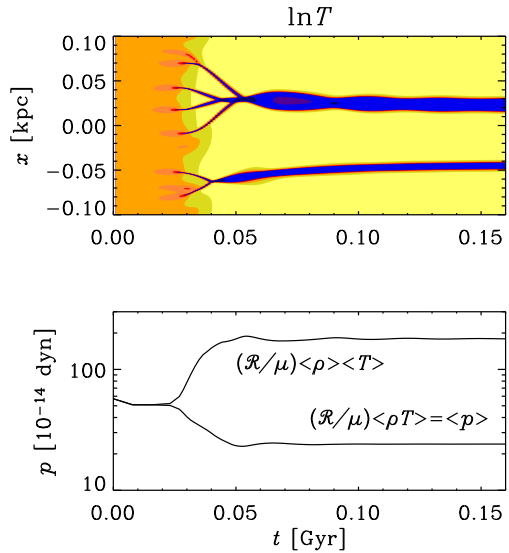}
\end{center}\caption[]{
Evolution of $\ln T$ in a space-time diagram (top) and of
the mean pressure (bottom) in a one-dimensional simulation with
$1024$ mesh points and $\nu=\chi=5\times10^{-4}\Gyr\,\km^2\s^{-2}$.
During early times, the rms velocity grows exponentially at
a rate of about $220\Gyr^{-1}$.
}\label{prho}\end{figure}

When setting up a simulation the density is particularly useful,
because its mean value in a certain volume is proportional to the
mass, which is constant for closed and periodic boundary conditions,
such as those considered here.
Thus, one can ask the question what is the resulting mean pressure
as a function of the mean density.
Of course, as long as the gas is thermally stable, the density
will be uniform and hence its mean value is always equal to the
actual value at any point, so it is given by combining the equation
of state with the condition of thermal equilibrium.
As is evident from \Fig{ppcoolheati}, when the density is in the range
$(1$--$5)\times10^{-24}\g\cm^3$, there is no stable solution.
This means that the gas will fragment into cold patches of temperature
$T=T_{\rm C}$ with density $\rho_{\rm C}$, and the rest of the ambient
gas warms up to the stable solution branch $T=T_{\rm W}$ with density
$\rho_{\rm W}$.
As a direct result of mass conservation in our periodic domain, the filling 
factor of the cold component can be expressed in terms of
the mean density, $\bra{\rho}$, which is known from the initial
condition.
Using the definition of the filling factor,
\EQ
f\rho_{\rm C}+(1-f)\rho_{\rm W}=\bra{\rho},
\EN
the value of $f$ is given by
\EQ
f={\bra{\rho}-\rho_{\rm W}\over\rho_{\rm C}-\rho_{\rm W}}.
\label{FillingFactor}
\EN
A similar analysis can also be adopted for calculating $\bra{T}$.
This allows us to calculate the correlation coefficient $\epsilon$
in the relation
\EQ
\bra{\rho T}=\epsilon\,\bra{\rho}\bra{T}
\EN
where
\EQ
\epsilon\approx{1\over f(1-f)}{T_{\rm W}\over T_{\rm C}}
\approx{0.013\over f(1-f)}.
\EN
The correlation coefficient is small because $\bra{\rho T}$ decreases
slightly and $\bra{\rho}\bra{T}$ increases strongly as the system
segregates, as demonstrated below (in connection with \Fig{prho}).
The expression $\bra{\rho}\bra{T}$ is almost entirely determined by the
product of the volume-weighted density (or relative mass) in the cold phase,
$f\rho_{\rm C}$, and the volume weighted temperature in the warm phase,
$(1-f)T_{\rm W}$, so both factors are large compared with their respective
average values.

The segregation phenomenon has already been studied in a one-dimensional
model (S{\'a}nchez-Salcedo et al.\ 2002).
Here, except for an additional perturbation,
the initial condition is assumed uniform, $\rho=\rho_0\equiv\bra{\rho}$,
and the value of $\rho_0$ is varied between different simulations.
In all the runs presented below the initial perturbation is gaussian
noise with an rms fluctuation amplitude of $10^{-26}\g\cm^{-3}$.
When $\rho_0$ (in units of $10^{-24}\g\cm^{-3}$) is between 0.96 and 5.1,
the gas is thermally unstable and segregates into cold and warm components.
As time goes on, some of the cold spots may move because of slight
pressure imbalance until they coalesce into bigger fragments.
This coalescence was also found by S{\'a}nchez-Salcedo et al.\ (2002)
and Koyama \& Inutsuka (2004).

In \Fig{prho} we plot the evolution of $\ln T$ in a space-time diagram
(top) and that of the mean pressure in a one-dimensional simulation.
Here, $\nu=\chi=5\times10^{-4}\Gyr\,\km^2\s^{-2}$ which, together with
the initial values of $c_{\rm s}=7.5\kms$ and $n_\rho=980\Gyr^{-1}$,
yields $k_{\rm F}=720\kpc^{-1}=23k_1$, and hence
$n_\rho/(c_{\rm s}k_{\rm F})\approx0.2$.

During early times the rms velocity grows exponentially at
a rate of about $210\Gyr^{-1}$, which is consistent with
the peak value of $n/(c_{\rm s}k_{\rm F})\approx0.04$ for our
set of parameters.
Note that the mean pressure settles around $24\times10^{-14}\dyn$ once
the instability has saturated.
At that time, smaller structures may still coalesce into larger ones,
but the total filling factor remains approximately constant.
During the evolution away from the unstable homogeneous state,
the mean pressure (proportional to $\bra{\rho T}$) decreases
by about a factor of 2, but the product $\bra{\rho}\bra{T}$ increases
by almost a factor of 4.
When $\rho_0$ is between 5.2 and 11 (in units of $10^{-24}\g\cm^{-3}$)
the gas is marginally stable ($\beta=1$; see \Tab{Tcooling}), so in that
range there will be no segregation into different phases.

\begin{figure}[t!]\begin{center}
\plotone{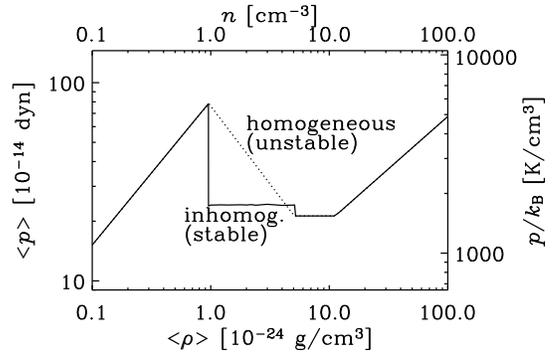}
\end{center}\caption[]{
Mean pressure vs.\ mean density in a one-dimensional simulation
(solid line), compared with the values obtained for a homogeneous
system (dotted line).
The dotted line (which agrees with the solid line in the stable regime
and hence cannot be seen there) was obtained by plotting $p(T)$ vs.\
$\rho(T)$ using $T$ as a parameter in \Eq{parametric}.
On the right axis, the pressure is normalized by the Boltzmann constant,
$p/k_{\rm B}$, and at the top the number density is given.
The simulation has $128$ mesh points and
$\nu=\chi=5\times10^{-3}\Gyr\,\km^2\s^{-2}$.
}\label{press1d}\end{figure}

When the mean density is outside the range between 0.96 and 5.1
(in units of $10^{-24}\g\cm^{-3}$), the gas is thermally stable
and remains uniform.
The dependence of the pressure on the density can be obtained in
parametric form by calculating, using temperature as a parameter,
$\rho(T)$ and $p(T)$, that is,
\EQ
\rho(T)={\Gamma\over\Lambda(T)},\quad
p(T)={{\cal R}T\over\mu}\,{\Gamma\over\Lambda(T)},
\label{parametric}
\EN
and plotting the two against each other (see \Fig{press1d}, dotted line).
The numerically obtained values for the mean pressure $\bra{p}$,
for different mean densities $\bra{\rho}$, agree with those obtained
under the assumption of homogeneity.

\begin{figure}[t!]\begin{center}
\plotone{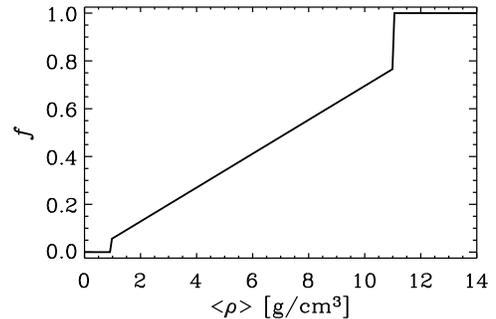}
\end{center}\caption[]{
Filling factor $f$ as a function of the mean density,
as predicted by \Eq{FillingFactor}.
}\label{pfilling}\end{figure}

In the unstable regime the pressure is, surprisingly,
independent of $\bra{\rho}$, and
always around $\bra{p}\approx24.2\times10^{-14}\dyn$.
(\FFig{press1d} shows slight variations about this value; this is
probably a consequence of the fact that the coefficients $C_{i,i+1}$
are only implemented up to three to four significant figures, so the cooling
curve is still not perfectly continuous.)
\FFig{ppcoolheati} shows that for $\bra{p}\approx24.2\times10^{-14}\dyn$
the warm and cool phases have $\rho_{\rm W}\approx0.19$ and
$\rho_{\rm C}\approx14.3$, respectively.
This allows us to determine the filling factor as a function of $\bra{\rho}$;
see \Fig{pfilling}.
In most of the runs considered below we expect
$\bra{\rho}=2$, so $f\approx13\%$.
In practice we estimate the filling factor as the fraction of mesh points
for which $T<T_{\rm U}$, where $T_{\rm U}\approx420\K$ (corresponding
to $\rho_{\rm U}=4.3\times10^{-24}\g\cm^{-3}$ in \Fig{ppcoolheati})
for $\bra{p}=24.2\times10^{-14}\dyn$.
The filling factors determined in this way are quoted for the simulations
presented below.

There is a tendency for cool patches to travel and to coalesce into
bigger ones (see, e.g., S{\'a}nchez-Salcedo et al.\ 2002).
This property is reminiscent of earlier work in the context
of the thermal instability.
Elphick et al.\ (1991, 1992) found traveling front solutions and also the
merging of smaller patches into bigger ones, which they associate
loosely with an inverse cascade behavior.
However, in their work they only discuss the energy equation and not
dynamical processes.
In the case they discuss the kink and antikink fronts always travel toward
or away from each other, thus resulting in the annihilation and creation
of denser clouds.
This is not seen in the present work.
Also, they discuss much smaller objects of size $\sim0.02\pc$ which have
considerably shorter sound crossing times.
Furthermore, early on in their evolution our clouds tend to accelerate
toward each other, as can be seen from the curved trajectories.

\section{Three-dimensional simulations}

\subsection{Fully periodic boundary conditions}

\begin{figure}[t!]\begin{center}
\plotone{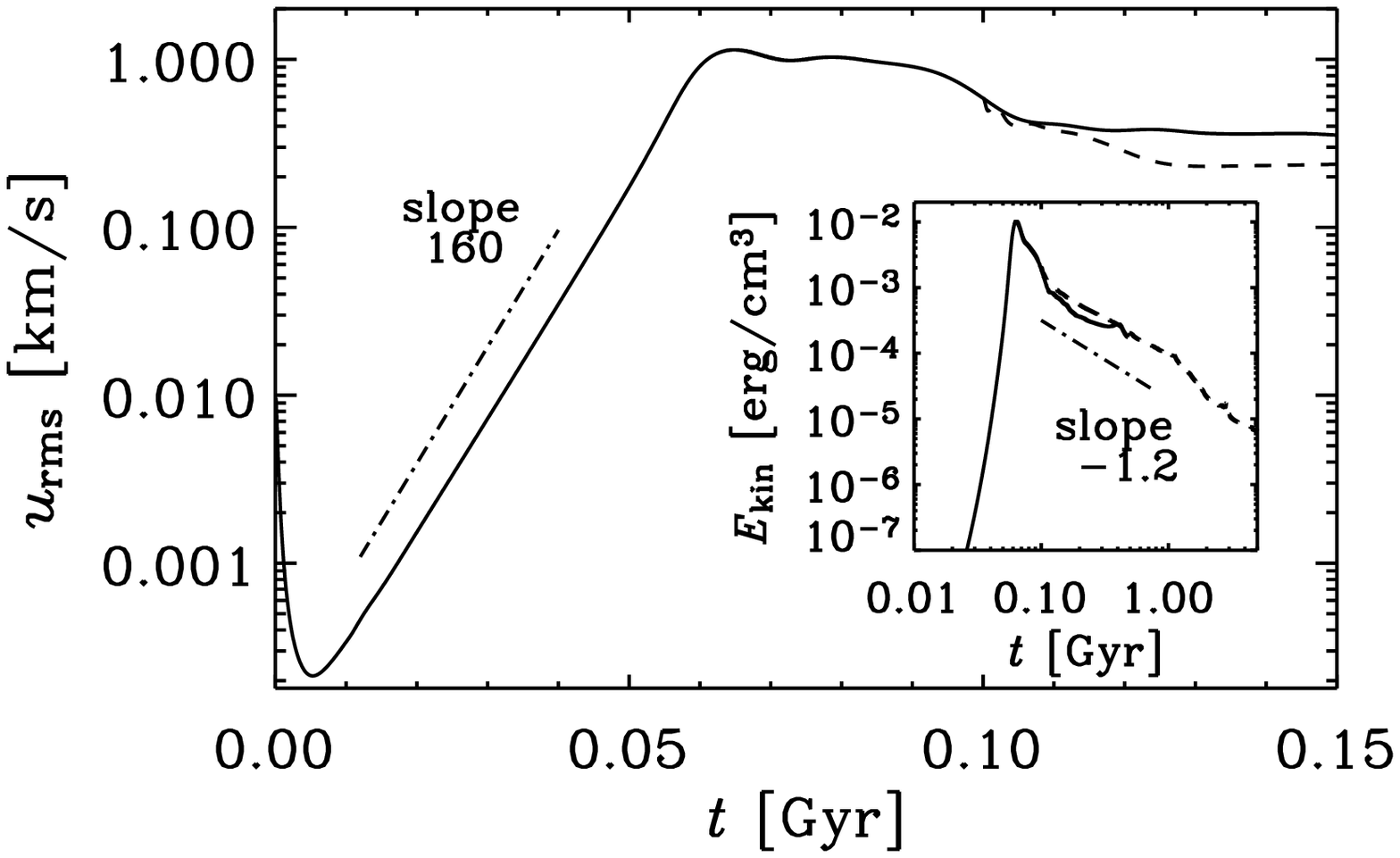}
\end{center}\caption[]{
The rms velocity and kinetic energy for a three-dimensional
run with $\nu=\chi=5\times10^{-3}\Gyr\,\km^2\s^{-2}$ and $256^3$ mesh points.
The dashed line represents a solution with
$\nu=\chi=5\times10^{-4}\Gyr\,\km^2\s^{-2}$, which was restarted
at $t=0.1\Gyr$ from a run with 10 times higher viscosity.
Here $\bra{\rho}\approx1.7\times10^{-24}\g\cm^{-3}$,
$\bra{p}\approx24.3\times10^{-14}\dyn$, and $\bra{T}\approx8200\K$.
}\label{purms}\end{figure}

In this section we discuss the results of three-dimensional simulations.
The basic properties of the one-dimensional simulations, presented in
\Sec{SaturationProperties}, carry over to the three-dimensional regime.
As expected, the growth rates are the same as
those found in the one-dimensional case.
The resulting mean pressure $\bra{p}$ and hence the filling factor,
as given by \Eq{FillingFactor},
are also quite similar to those of the one-dimensional case.
Nevertheless, even though significant amounts of turbulent heating are
being produced at the most violent phase of the instability, there is
in our cases always a subsequent relaxation phase in which the flow speed
tends to vanish on a long time scale (see \Fig{purms}).
This agrees with earlier findings of Kritsuk \& Norman (2002a).
The energy decay is consistent with a $t^{-1.2}$ law, just like in ordinary
turbulence (e.g.\ Mac Low et al.\ 1998, Haugen \& Brandenburg 2004).
This is also consistent with the results of Kritsuk \& Norman (2002a),
who reported decay exponents in the range 1--2 for box sizes between 5 and
$500\pc$ using also a more detailed cooling curve in tabular form.
On the other hand, Koyama \& Inutsuka (2006) find that turbulence remains
self-sustained for times up to $0.1\Gyr$.

We emphasize again that we have used constant kinematic viscosity and
constant thermal diffusivity in our simulations.
For the runs shown in \Figss{purms}{ppdf2d} we have used
$\nu=\chi=5\times10^{-3}\Gyr\,\km^2\s^{-2}$ until $t=0.1\Gyr$
(corresponding to $\mbox{Re}_{\rm mesh}=2$).
This corresponds to $k_{\rm F}=230\kpc^{-1}=23k_1$, and hence
$n_\rho/(c_{\rm s}k_{\rm F})\approx0.6$, so the initial growth
rate is $160\Gyr^{-1}$.
This is again consistent with \Fig{prho} yielding a peak value of
$n/(c_{\rm s}k_{\rm F})\approx0.09$ for our set of parameters.

\begin{figure*}\centering
\plotone{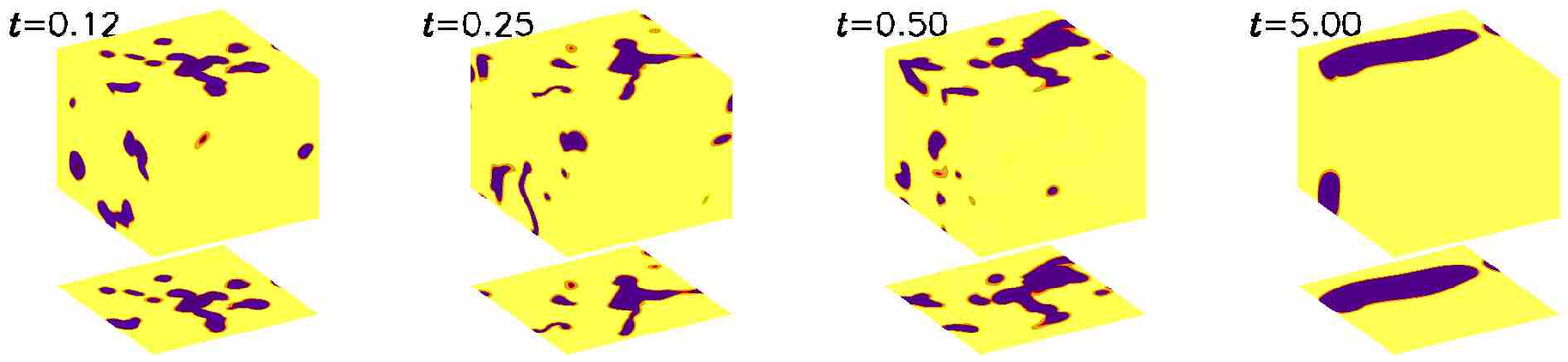}
\caption{
Visualization of $\ln T$ on the periphery of the box at different times,
for $\nu=\chi=5\times10^{-4}\Gyr\,\km^2\s^{-2}$ with $256^3$ mesh points.
$\bra{\rho}\approx1.7\times10^{-24}\g\cm^{-3}$,
$\bra{p}\approx24.3\times10^{-14}\dyn$, $\bra{T}\approx8200\K$, and $f=11\%$.
For this run $k_{\rm F}/k_1=130$ and $n_\rho/(c_{\rm s}k_{\rm F})=0.37$.
Prior to $t=0.1\Gyr$, both viscosity and thermal diffusivity were 10 times
larger; $\nu=\chi=5\times10^{-3}\Gyr\,\km^2\s^{-2}$ with
$k_{\rm F}/k_1=41$ and $n_\rho/(c_{\rm s}k_{\rm F})=1.2$.
The growth rate is about $160\Gyr^{-1}$.
Note the isolated cool patches (dark shaded) compared with the extended
warm background (light shades).
As time goes on, the dark patches merge with each other and grow.
}\label{img}\end{figure*}

However, after having reached the peak velocity, the flow has becomes
sufficiently quiescent so that it is possible to decrease the viscosity
by a factor of about 10, corresponding to $\mbox{Re}_{\rm mesh}=20$.
\FFig{img} shows images of $\ln T$ on the periphery of the simulation domain at
a few selected times after having lowered the viscosity and thermal viscosity.
Animations of temperature and density\footnote{\url{see
http://www.nordita.dk/~brandenb/movies/thermal\_inst}.}
show that late in the simulation, cold patches of gas are still moving
about, but this is presumably just a response to small-amplitude, small 
wavenumber variations in overall pressure requiring a much longer 
time scale to equilibrate.

During the course of the simulation the value of $k_{\rm F}$
(based on the averaged value of $n_\rho$) increases
between the initial value before saturation of the instability
($k_{\rm F}\delta x\approx0.2$) and the saturated state
($k_{\rm F}\delta x\approx1$ with the higher viscosity and
($k_{\rm F}\delta x\approx3$ with the lower viscosity).
At the end of the simulation the gas is sharply segregated into warm and
cool phases in almost perfect pressure equilibrium.
This can be seen clearly from probability density functions of the
various quantities that are discussed in \Sec{ForcedSimulations}.

\subsection{Shearing-periodic boundary conditions}

The shearing sheet approximation simulates the local conditions in a
disk with strong radial differential rotation in the limit of large radii.
Curvature can thus be neglected and the shear can be assumed linear in
radius, so that we only have an underlying linear shear flow
$\UU_0=(0,Sx,0)^T$, where $S$ is constant.
The Coriolis force, $2\OO\times\uu$ is added, where $\OO=(0,0,\Omega)$
is the angular velocity vector.
It is assumed that $S$ scales with the angular velocity, so here
we take $S=-\Omega$ which is appropriate for galactic disks
with a constant linear velocity law.
The combined effects of shear and Coriolis force can be subsumed into
a single vector (Brandenburg et al.\ 1995),
\EQ
\ff(\uu)=\pmatrix{2\Omega u_y\cr-(2\Omega+S) u_x\cr0},
\label{ffuuDefn}
\EN
which is then added on the right hand side of \Eq{dudt}.
After this modification the velocity $u$ describes the deviation
from the shear flow and does thus not include the basic shear.
The basic shear flow appears still explicitly as an additional advection
operator of the form $\UU_0\cdot\nab=Sx\partial_y$.

\begin{figure*}\centering
\plotone{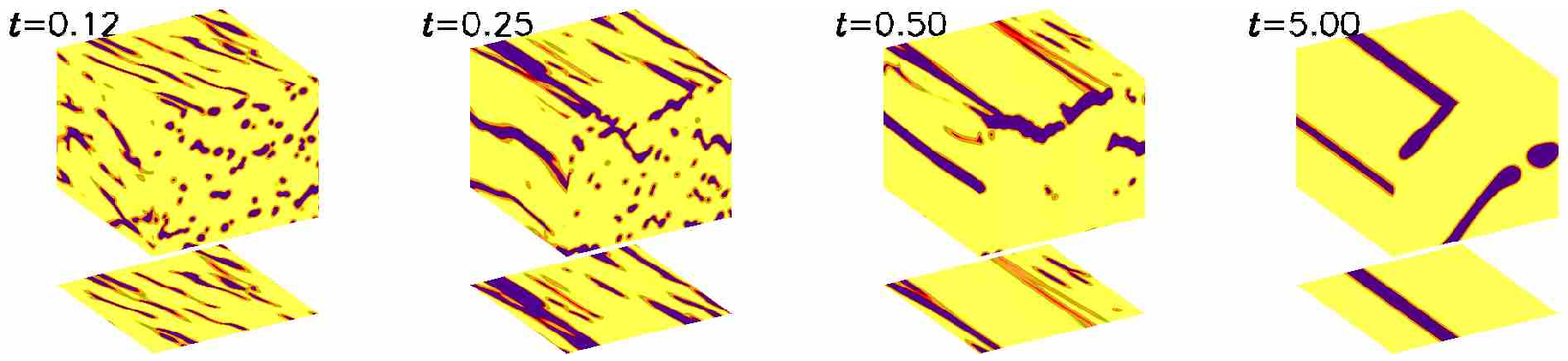}
\caption{
Visualization of $\ln T$ on the periphery of the box at different times,
for $\nu=\chi=5\times10^{-4}\Gyr\,\km^2\s^{-2}$ and $256^3$ mesh points.
$\bra{\rho}\approx1.74\times10^{-24}\g\cm^{-3}$,
$\bra{p}\approx24.2\times10^{-14}\dyn$, and $\bra{T}\approx8200\K$.
Here $\Omega=100\Gyr^{-1}$ and $S=-\Omega$.
For this run $k_{\rm F}/k_1=32$ and $n_\rho/(c_{\rm s}k_{\rm F})=1.5$.
The growth rate is about $190\Gyr^{-1}$, which is somewhat larger
than for the corresponding non-shearing run.
Note that the initially produced structures are quickly sheared out.
}\label{img2}\end{figure*}

In the following we consider the case $\Omega=100\Gyr^{-1}$, but we
have also considered the case $\Omega=25\Gyr^{-1}$ (appropriate for
our Galaxy).
The difference between the two simulations is small.
The main thing that happens in all these simulations is a tendency for
the flow to become sheared out, so any variations in the streamwise
direction become sheared out and the flow becomes essentially two-dimensional;
see \Fig{img2}.
However, shear does not seem to lead to instability, even though the
kinematic growth rate of the thermal instability is apparently somewhat
increased ($190\Gyr^{-1}$ instead of $160\Gyr^{-1}$).
This absence of sustained turbulence is somewhat disappointing,
because one might have hoped that the
thermal instability would have led to condensation in the streamwise
direction and thus to new structures that could then be sheared out again.
This seems to be prevented by the general tendency of coalescence,
preventing breakup into new structures in the streamwise direction.
However, it may still be interesting to reconsider this issue in the future
at significantly higher resolution and larger Reynolds number.

\subsection{Forced simulations}
\label{ForcedSimulations}

Given that the TI did not produce sustained turbulent flows, we consider
now cases which turbulence is driven by an additional body force in
the momentum equation.
We consider here a forcing function consisting of plane waves
whose wavevector is chosen randomly at each time step and has length
between 2.5 and 3.5 times the smallest wavenumber in the box,
$k_1=2\pi/(0.2\kpc)$.
This forcing function is therefore $\delta$-correlated in time and
approximately monochromatic in space (see also
S{\'a}nchez-Salcedo et al.\ [2002] for simulations in one dimension).

\begin{figure}[t!]\begin{center}
\plotone{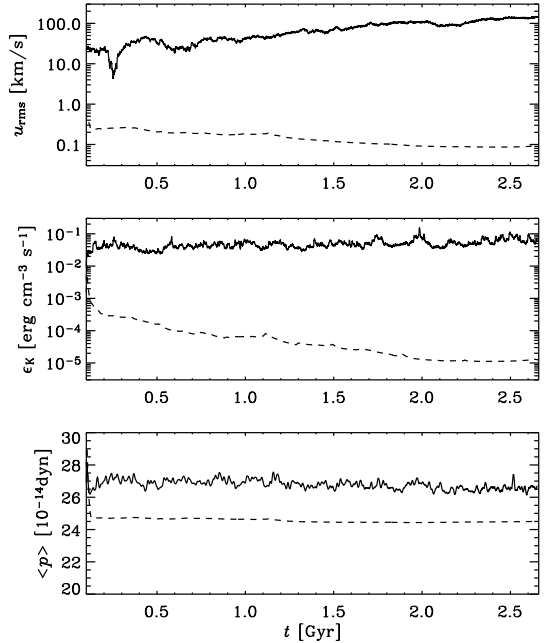}
\end{center}\caption[]{
Comparison of the rms velocity (top), dissipation rate
$\epsilon_{\rm K}$ (middle), and mean pressure (bottom),
for a forced simulation (solid lines) and a nonforced
simulation (dotted lines; same run as in \Fig{img}).
Both cases have $\nu=\chi=5\times10^{-4}\Gyr\,\km^2\s^{-2}$ and
$256^3$ mesh points.
}\label{purms_comp}\end{figure}

\begin{figure}[t!]\begin{center}
\plotone{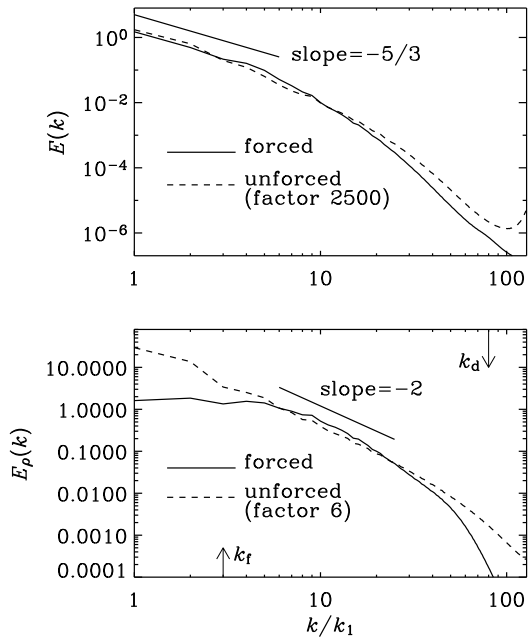}
\end{center}\caption[]{
Comparison of time-averaged kinetic energy spectra (top)
and density spectra (bottom) for the forced and unforced runs
shown in \Fig{purms_comp}.
The kinetic energy and density spectra for the unforced case are
scaled so as to make them overlap at intermediate wavenumbers.
For the forced run the forcing wavenumber, $k_{\rm f}=2.5k_1$, is
indicated, while the dissipation wavenumber,
$k_{\rm d}=(\omega_{\rm rms}/\nu)^{1/2}$, is shown for the unforced run.
}\label{pspec_comp}\end{figure}

It turns out that when the flow is driven sufficiently strongly to
produce rms velocities of around $10$--$30\kms$, the turbulent energy that
is dissipated into heat is only about comparable to the energy
needed to balance the losses from cooling (see \Fig{purms_comp}).
The mean pressure is increased slightly to about $30\times10^{-14}\dyn$,
corresponding to $p/k_{\rm B}\approx2170\K\cm^{-3}$.
In both cases the spectra of $\uu$ (kinetic energy) and $\rho$ (density)
are similar, except that the unforced run shows more relative power in
the density spectra at large scales; see \Fig{pspec_comp}.
Over a small range of wavenumbers the local slope of the kinetic energy
spectra is around $-5/3$.
By comparison, Kritsuk \& Norman (2004) found shallower spectra with
spectral slope close to $-1$ in their decaying simulations with TI, but
this could be a feature of the numerical dissipation used in their code.
The dissipation wavenumber, $k_{\rm d}=(\omega_{\rm rms}/\nu)^{1/2}$,
where $\omega_{\rm rms}$ is the rms of the vorticity, $\oo=\nab\times\uu$
is shown for the unforced run.
For the forced run the adopted viscosity is critically low, as
evidenced by the small rise in the kinetic energy at large wavenumbers.
In fact, the dissipation scale for this run is just outside the plot range.
It is perhaps because of the presence of cooling, which contributes to
energy removal, that this run has still been successful.

\begin{figure}[t!]\begin{center}
\plotone{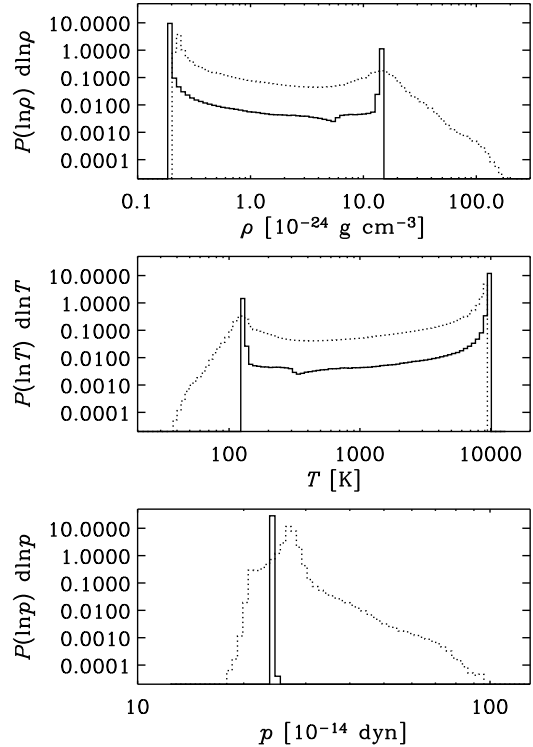}
\end{center}\caption[]{
Comparison of the
probability density functions of $\ln\rho$, $\ln T$, and $\ln p$ at the last
snapshot for forced (dotted lines) and unforced (solid lines) cases. 
}\label{ppdf}\end{figure}

As in the unforced case, the gas is segregated into warm and
cool phases, but now they are only in {\it approximate} pressure equilibrium;
in \Fig{ppdf}, we show probability density functions (PDFs)
of $\ln\rho$, $\ln T$, and $\ln p$. 
The turbulent increase of the
mean pressure has only a small effect on the preferred
temperatures in the warm and cold phases, whereas the density peaks
are shifted toward higher densities, as expected if the system were
still following the equilibrium pressure-density relation. The
turbulence forced at relatively small scales has the most drastic
effect on the cold cloudy component, the distribution of which has
become significantly wider while the high density wing was developing.
The maximum density in the forced case is roughly an order of magnitude
larger than in the pure TI case.
A similar wing is observed at low temperatures, reaching values down to
the cooling cut-off of $10\K$ in the highest density regions.
While in the
pure TI case the pressure in the saturated state shows a very narrow
distribution around the mean, in the forced cases the distribution is
broad with extrema that vary by almost 1 order of magnitude.
In addition to this broadening of the pressure distribution
already pointed out in several previous studies (e.g.\ Gazol et
al.\ 2005), the amount of gas in the ``forbidden'' (thermally unstable)
regime has been observed to increase; in our calculations this is
seen as a systematic increase of the level of the PDFs in between the
two preferred states, while the peaks themselves become less
pronounced.
In the forced case, about 6\% of the gas is found in the unstable range
where $\rho$ is between 1 and 5 times $10^{-24}\g\cm^{-3}$.
In the pure TI case, on the other hand, only 2\% is in this range.
In addition
there is a significant fraction of cold high-density overpressured gas
that is in the thermally stable regime.
Nevertheless, even in the
highly turbulent regime the signatures of pure TI are still
clearly visible in the density and temperature PDFs (better for warm,
worse for cold); the pressure PDF develops broad wings,
as is familiar from supernova-driven turbulence simulations
(e.g., Korpi et al.\ 1999; Mac Low et al.\ 2005), and some earlier
TI simulations with forced turbulence (e.g.\ Gazol et al.\ 2005). 
Still, the mean pressure determines the preferred densities and temperatures in
the warm and cold phases as though the system followed the equilibrium
pressure-density relation.
 
\begin{figure}[t!]\begin{center}
\plotone{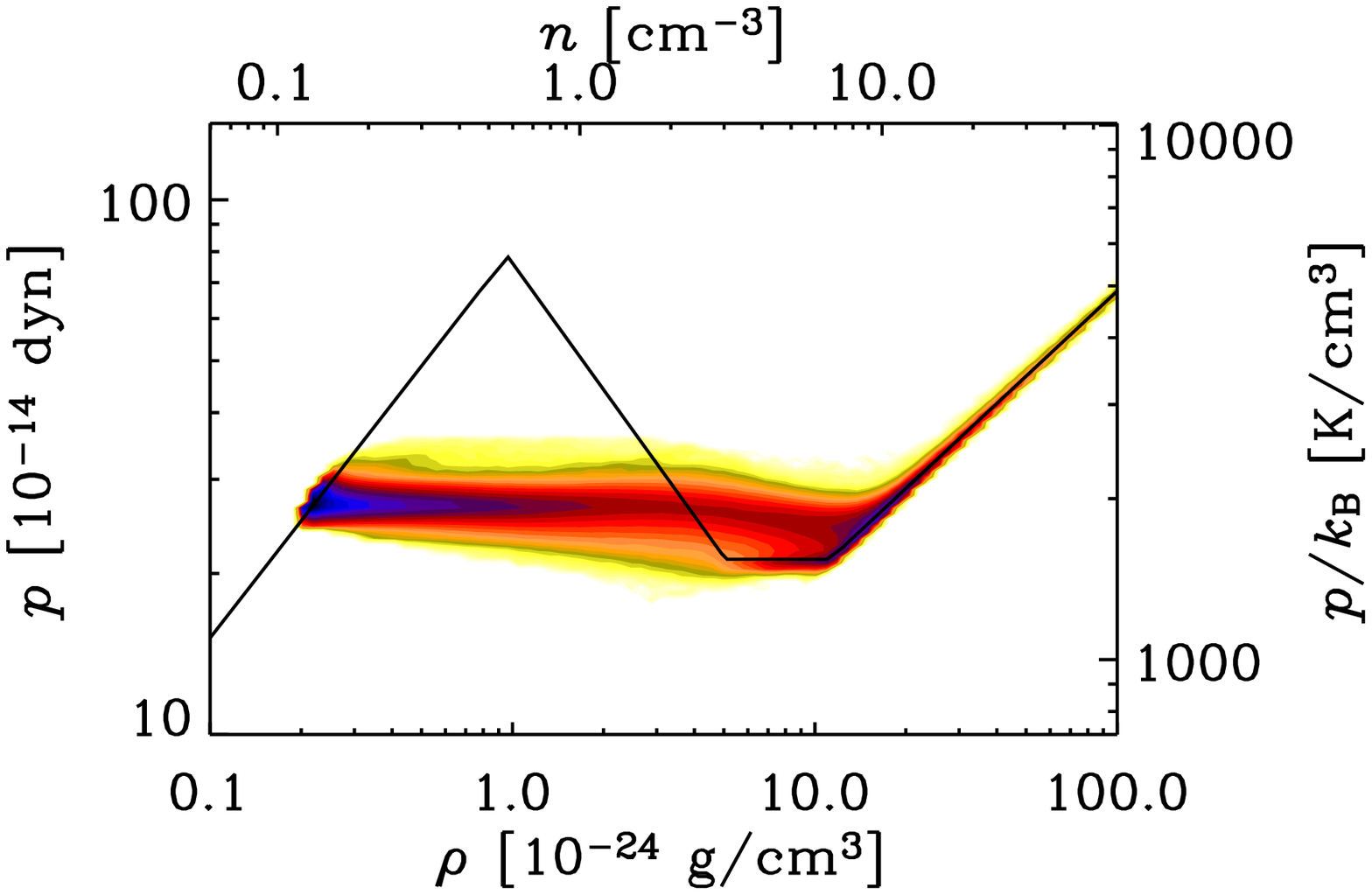}
\plotone{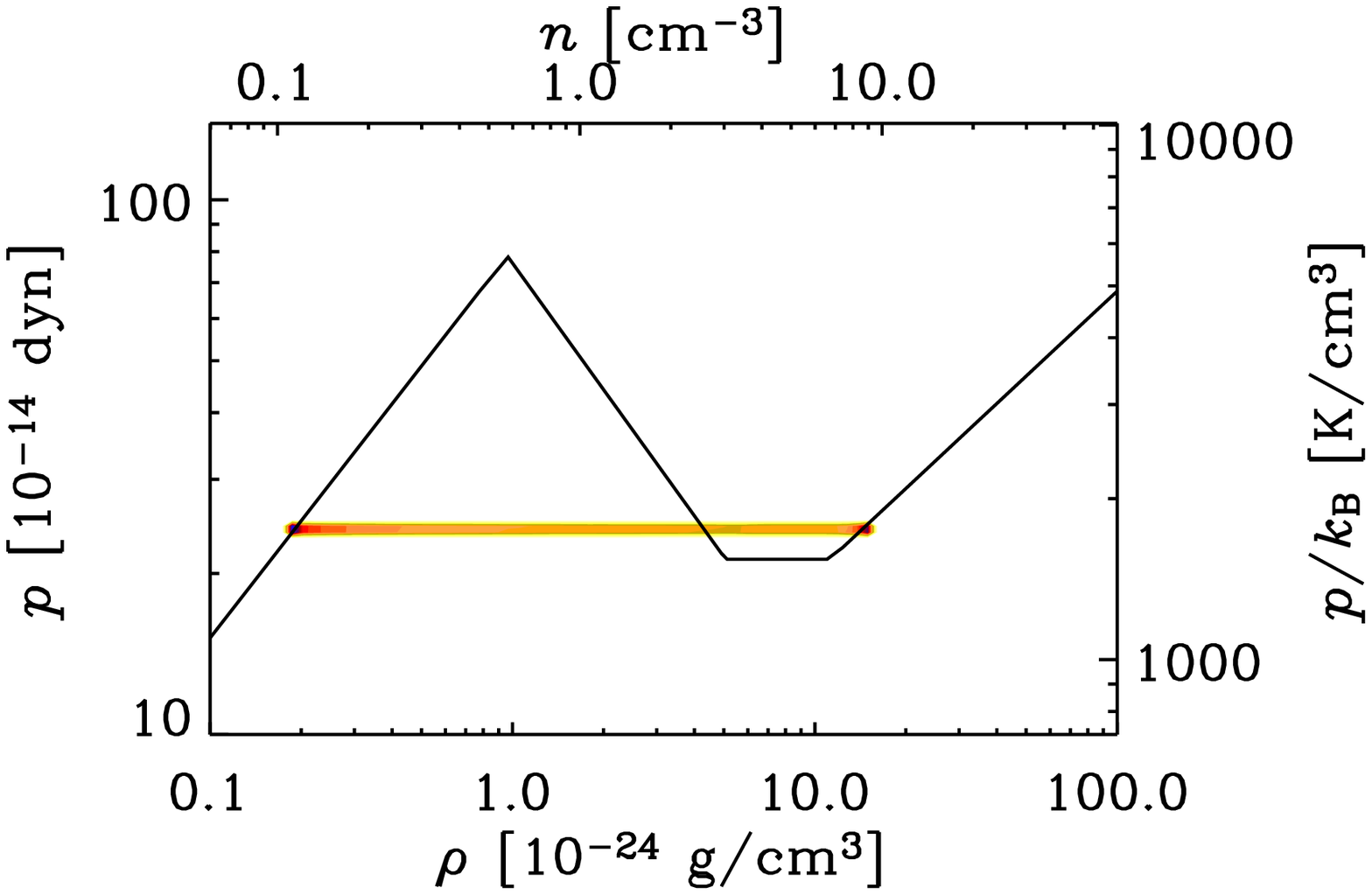}
\end{center}\caption[]{
Two-dimensional probability density functions of $\ln\rho$ and $\ln p$
for forced (top) and unforced (bottom) simulations.
The solid line indicates the thermodynamic equilibrium solution.
Dark shades indicate large values of the probability density.
}\label{ppdf2d}\end{figure}

It is customary to discuss scatter plots of pressure versus density
(S{\'a}nchez-Salcedo et al.\ 2002, Piontek \& Ostriker 2004, 2005), which
allow one to discuss the degree to which the gas is locally in equilibrium.
In \Fig{press1d} we showed that the mean pressure, that is, averaged over
the entire box, is $\approx24\times10^{-14}\dyn$ when the mean density is
in the unstable range, $\bra{\rho}=(1$--$5)\times10^{-24}\g\cm^{-3}$.
It turns out that this result also holds locally, as can be seen from
a scatter plot of pressure versus density and, more conveniently, from
a two-dimensional PDF (\Fig{ppdf2d}) showing the logarithm of the
probability density as a function of both $\ln\rho$ and $\ln p$ for
both forced and unforced runs.
In the unforced case the local pressure is concentrated sharply around
$24\times10^{-14}\dyn$ over a broad range of local densities,
$\rho=0.2$--$20\times10^{-24}\g\cm^{-3}$.
In the forced case, the distribution is broadened around the average
pressure for $\rho=0.2$--$20\times10^{-24}\g\cm^{-3}$, but there are also
dense spots with $\rho\ga20\times10^{-24}\g\cm^{-3}$ where the gas follows
the equilibrium distribution quite sharply, confirming earlier findings
of S{\'a}nchez-Salcedo et al.\ (2002) and Piontek \& Ostriker (2004, 2005).

\section{Conclusions}

Our results confirm the basic findings of Kritsuk \& Norman (2002a)
in that the TI does not lead to self-sustained turbulence.
In the cases considered
in this paper the instability just leads to segregation into two
different phases, and produces only small velocities in response to
the remaining pressure fluctuations. While the growth of the
instability occurs over relatively short timescales of a few tens of
millions of years, the kinetic energy of these motions decays exponentially
with a slope consistent with $-1.2$ leading to insignificant rms
velocities after a few hundred millions of years.
Thus, in agreement with Kritsuk \& Norman (2002a),
the TI alone does not lead to self-sustained turbulence. 
This is somewhat different from the purely two-dimensional TI cases 
investigated by Piontek \& Ostriker (2004), who report weak
($\approx 0.5\km\s^{-1}$) non-decaying turbulence over timescales of
roughly $0.5\Gyr$. This behavior seems to carry over into three
dimensions (Piontek \& Ostriker 2005, see their Fig.~11).

Similar results have also been found recently by Koyama \& Inutsuka (2006),
who also include an explicit dynamical viscosity.
For times up to $0.1\Gyr$ their results are nevertheless qualitatively
similar to ours in that they also report rms velocities in the range
$0.1$--$0.4\kms$, and their flow topology is similar to ours at early times.
They also study smaller box sizes, but their highest turbulence levels
occur for their largest box size of $L=144\pc$, which is similar to ours.
In both cases the Field length is about 1/100 of the box size.
However, if there is really a difference in sustaining turbulence
over long times, then this might be due to a different
formulation of thermal conduction which varies here with density,
but is constant in the simulations of Piontek \& Ostriker (2004, 2005)
and Koyama \& Inutsuka (2006).
In the latter case a constant dynamical viscosity is included, while
in our case a constant kinematic viscosity is used.
Another difference is the discontinuous nature of the transitions in the
previously used cooling function.
(The latter was observed to lead to spurious oscillatory motions in
some of our preliminary investigations that are not reported here.)
However, if the turbulence is real, then this could perhaps be understood as an
analogy to the TI-driven turbulence found by Kritsuk \& Norman (2002b)
in the presence of a time-dependent heating rate.
The idea would be that a variable heating rate could perhaps be simulated
by introducing nonlinear feedbacks in some of the coefficients.

Another possibility for driving turbulence has been discussed by
Murray et al.\ (1993).
They find that a system segregated into two phases by the TI could develop
Kelvin-Helmholtz secondary instabilities if cold clouds move at transonic
speeds relative to a warm background.
They speculate that such motions could be the result of buoyancy forces
or some pressure imbalance.
However, this scenario does not seem to apply to our simulations where
pressure imbalances become quite small at late times.
A related possibility would be secondary instabilities caused by
differential rotation.
Again, in the present simulations this did not occur either.
Instead, shear mainly causes the flow to become two-dimensional, that is,
uniform in the streamwise direction.
However, in the simulations the TI shows no tendency of
subsequent fragmentation of structures in the streamwise direction.
There might still be some hope that the fragments could be susceptible to
a baroclinic instability, but this may require substantially higher resolution
than what we have considered in the present paper.

In the pure TI, cases the system develops into a new segregated state
in which each phase is stable.  The cold patches have a tendency to
coalesce into bigger ones that are more resistant to the
possibility of breaking up.
It is conceivable that the process of coalescence is slowed down
when the value of $\chi$ is decreased.
This might become more plausible when realizing that, because of the
thermal instability, the energy equation is essentially of the type of
a reaction-diffusion equation.
Under the assumption of perfect pressure equilibrium at all times,
Elphick et al.\ (1991) showed that this equation permits traveling
kink solutions.
If a front were to travel into an unstable equilibrium state, the front
speed would be proportional to the square root of the product of
diffusivity and the growth rate of the instability.
In the present case, however, warm and cold equilibrium states
``compete'' against each other, so fronts would not propagate.
Only in two and three dimensions, where fronts are in general curved,
they tend to be driven diffusively toward the direction of the center of
curvature; see Shaviv \& Regev (1994) as well as
Brandenburg \& Multam\"aki (2004) for similar results in
a different context.
However, the assumption of perfect pressure equilibrium is problematic,
because then the density is assumed to be inversely proportional to the
temperature, so mass conservation is generally not obeyed.
In our cases there is no perfect pressure equilibrium and
one may argue that the coalescence is primarily the
result of individual dense spots moving with the flow toward local
and global pressure minima.

It is in principle possible that the amount of viscous
heating might suffice to heat the cold patches enough to make
them unstable again.  However, the amount of viscous heating is
insufficient in all the cases that we have investigated.  Only when an
external forcing function is added to give the flow a rms
velocity of $10$--$30\kms$ does the total amount of heating become
comparable to $\Gamma$, that is, the level of the imposed uniform heating.
Obviously, we cannot exclude the possibility of TI-driven turbulence
for smaller viscosity, smaller thermal diffusivity, or both.
It may therefore be useful to revisit this issue in future when simulations
at higher resolution become more affordable.

Detailed models of the heating and cooling properties of the warm and
cold components of the ISM have been used to calculate the equilibrium
curves, which in practice predict the range of stable versus unstable
densities, temperatures and pressures in the ISM (see, e.g., Wolfire et
al.\ 1995, 2003). Calculations, such as those presented in this
paper, of the onset and nonlinear stages of the TI, are needed in order to
investigate the actual equilibrium pressure realized in a system
described by a certain equilibrium curve; from this, the
characteristic densities, temperatures and filling factors for the
warm and cold phases can also be determined.

Our one-dimensional calculations (\Fig{press1d}) of the standard
model of Wolfire et al.\ (1995)
show that the mean pressure realized in the unstable regime remains
roughly constant at $1750\K\cm^{-3}$ over the whole range of unstable
densities, $\bra{\rho}=(1$--$5)\times10^{-24}\g\cm^{-3}$, and that the
pressure is close to the minimal value of $1540\K\cm^{-3}$.
We note, however, that the equilibrium curve differs from
the original one because we have used $\mu$=0.62 instead of $\mu$=1,
and it also differs somewhat from S{\'a}nchez-Salcedo et al.\ (2002) as we
have used revised coefficients based on more accurate
continuity considerations.
The corresponding
temperatures and number densities at this equilibrium pressure are
$T_{\rm C}=126\K$, $n_{\rm C}=8.6\cm^{-3}$ and $T_{\rm W}=9430\K$,
$n_{\rm W}=0.1\cm^{-3}$.  This behavior carries over into the
three-dimensional regime.  The calculations with forced turbulence,
where the strongest forcing results in turbulent pressures exceeding
the thermal pressure by a factor of $\la3$ show that the mean pressure
increases only by about 25\%, even though the level of turbulence
is relatively strong.
The mean pressure obtained in this case in the
three-dimensional calculations is roughly $2170\K\cm^{-3}$, which is
in agreement with the observed median pressure of $p/k_{\rm
B}\approx2250\K\cm^{-3}$ from Jenkins \& Tripp (2001). 

\acknowledgements
We thank the referee for detailed comments, which have motivated us to
consider higher resolution runs and helped us to improve
the paper.
MJK and AJM thank Nordita for hospitality during the course of this study.
This work was supported by the Academy of Finland through grant
1112020 (MJK) and the Particle Physics and Astronomy Research Council
through grant PPA/S/S/2002/03473 (AJM).
The Danish Center for Scientific Computing is
acknowledged for granting time on the Horseshoe and Steno clusters.


\end{document}